\documentclass[twocolumn,showpacs,prl]{revtex4}

\usepackage{graphicx}%
\usepackage{dcolumn}
\usepackage{amsmath}
\usepackage{latexsym}

\begin {document}

\title
{
Modulated Scale-free Network in the Euclidean Space                                          
}
\author
{
S. S. Manna$^1$ and Parongama Sen$^2$
}
\affiliation
{
$^1$Satyendra Nath Bose National Centre for Basic Sciences
    Block-JD, Sector-III, Salt Lake, Kolkata-700098, India \\
$^2$Department of Physics, University of Calcutta, 
    92 Acharya Prafulla Chandra Road, Kolkata 700009, India. 
}
\begin{abstract}

       A random network is grown by introducing at unit rate randomly 
   selected nodes on the Euclidean space. A node is randomly connected to 
   its $i$-th predecessor of degree $k_i$ with a directed link of 
   length $\ell$ using a probability proportional to $k_i \ell^{\alpha}$. Our numerical 
   study indicates that the network is Scale-free for all values of 
   $\alpha > \alpha_c $ and the degree distribution 
   decays stretched exponentially for the other values of $\alpha$. The link length
   distribution follows a power law: $D(\ell) \sim \ell^{\delta}$ 
   where $\delta$ is calculated exactly for the whole range of values of $\alpha$.

\end{abstract}
\pacs {05.10.-a, 05.40.-a, 05.50.+q, 87.18.Sn}
\maketitle


\eject

      Statistical properties of many different networks are being studied 
   recently with much interests. Examples include the World-wide web (WWW)
   \cite {web}, the internet structure \cite {Faloutsos}, neural networks 
   \cite {neural}, collaboration network \cite {collab} etc. Broadly these 
   networks are classified into four different groups, namely, (i) Networks
   on regular lattices (ii) Random networks \cite {random} (iii) Small-world 
   networks (SWN) \cite {WS} and (iv) Scale-free networks (SFN) \cite 
   {barabasi,albert}.
   
      It has been observed that the degree distributions 
   of nodes for two very important networks e.g., World-wide web \cite {web} which is 
   a network of webpages (nodes) and the hyperlinks (links) among various pages and the 
   internet network \cite {Faloutsos} of routers or autonomous systems follow power law as: 
\begin {equation}
P(k) \sim k^{-\gamma}.
\end {equation}
   These networks are called Scale-free networks and the exponent $\gamma$
   varies between 2 and 3 for these networks.
   Barab\'asi and Albert (BA) proposed a simple model for an evolving SFN that has the following two 
   essential ingredients, 
   namely: (i) A network grows from an initial set of $m_o$ nodes with 
   $m < m_o$ links among them. Further, at every time step a new node is introduced 
   and is randomly connected to $m$ previous nodes. (ii) Any of these $m$ links
   of the new node introduced at time $t$ connects a
   previous node $i$ with an attachment probability $\pi_i(t)$ which is linearly
   proportional to the degree $k_i(t)$ of the $i$-th node at time $t$
\begin {equation}
\pi^{BA}_i(t) \sim k_i(t).
\end {equation}
For BA model $\gamma=3$ \cite {albert}.

      The physical distance, or the Euclidean distance between the nodes plays
   an important role in cases like electrical networks, internet or even in
   postal and  transports networks etc. In these networks one tries to minimize the lengths of
   the connections e.g., electrical wires, ethernet cables or say travel
   distances of postal carriers.
   Static networks, in which connection probabilities depend on the Euclidean
   distance have already been considered in the context of small-world properties
   \cite{euclid}. Recently a number of interesting works in the context of SFN have
   been published \cite {jost,dorog}.

      Study of internet's topological structure is important for designing
   efficient routing protocols and modelling internet traffic. Waxman model
   describes internet with exponentially decaying link length distribution:
   $D(\ell) \sim exp(-\ell/\ell_o)$ \cite {Waxman}. Faloutsos et. al.
   observed the scale-free degree distribution of the internet \cite{Faloutsos}.
   Yook et. al. observed nodes of the router level network maps of North America
   are distributed on a fractal
   set and the link length distribution is inversely proportional to the link
   lengths \cite {yook1}. They also argued that a competition exists beteen 
   the preferential attachment of the nodes and the weightage of the link 
   lengths \cite {yook1}.

      In this paper our aim is to study how a
   scale-free network defined on the Euclidean space behaves when the usual BA
   attachment probability as in eqn. (2) is modulated by a link length $\ell$
   dependent factor $\ell^{\alpha}$. Our important observation is even for the
   uniform random distribution of nodes we obtain the power law variation of
   the link length distribution for all values of $\alpha$ including the
   empirically observed inverse variation where as the Waxman's exponential
   behaviour is never observed. We argue that for a country with homogeneously
   distributed router density, our results seem to be important.

      Specifically in two-dimensions, we consider an unit square area on the $x-y$ plane.
   Randomly selected points within this area are the nodes of the network. The
   network grows by systematically introducing one node at a time with randomly 
   chosen coordinates $(x,y); 0\le x,y<1$ with uniform probabilities.
   The attachment probability that the new node introduced at time $t$ would 
   be connected to its $i$-th predecessor $(0 \le i \le t-1)$ is:
\begin {equation}
\pi_i(t) \sim k_i(t) \ell^{\alpha}
\end {equation}
   where, $\ell$ is the Euclidean distance between the $t$-th and the $i$-th node
   and $\alpha$ is a continuously varying parameter.

      The case with $\alpha$=0 is the usual BA model. 
   For the negative values of $\alpha$, the largest value of the modulation 
   factor $\ell^{\alpha}$ corresponds to the smallest value of $\ell$. Therefore, 
   in the limit of $\alpha \rightarrow -\infty$, only the smallest value of $\ell$ 
   corresponding to the nearest node will contribute with probability one. 
   Similarly, for $\alpha > 0$ large $\ell$ values will be more probable and the 
   limit of $\alpha \rightarrow +\infty$ corresponds to only non-zero contribution 
   from the furthest node. 
   
\begin{figure}[t]
\begin{center}
\includegraphics[width=8cm]{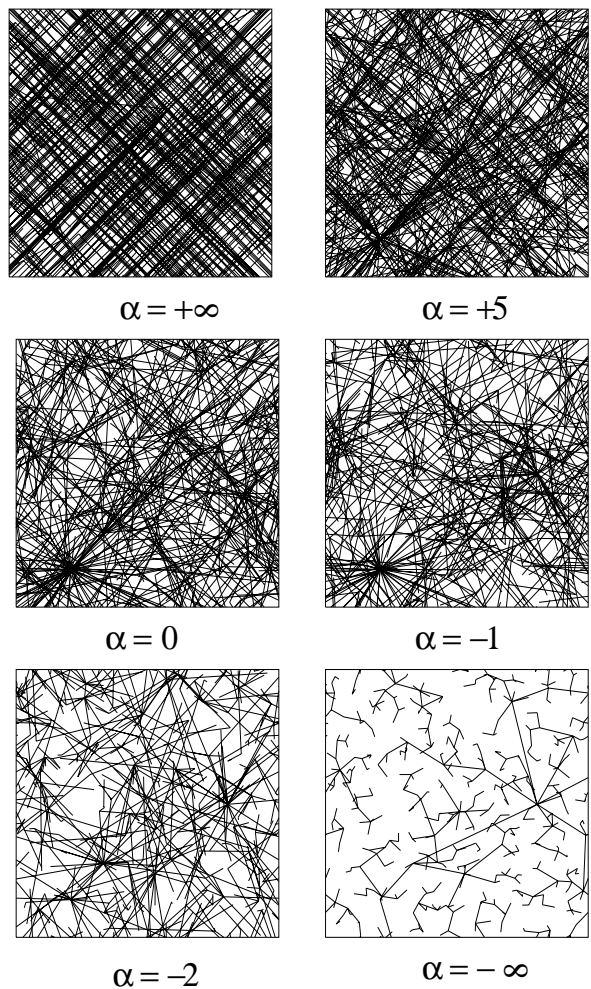}
\end{center}
\caption{
Modulated scale free networks within an unit square for different values
of the modulation parameter $\alpha$ for the same distribution of 512 nodes. 
For $\alpha=+\infty$ and $-\infty$ a newly introduced node is linked only to its 
farthest and nearest predecessors respectively, where as for $\alpha=0$ it 
is connected to one of the previous node according to the BA rule.
}
\end{figure}

      We start with only one initial node corresponding to $m = m_o = 1$ and 
   connect a new node to only one of its previous nodes. Therefore, every
   node has only one outgoing link $(k_{out}=1)$ but can have a number of 
   incoming links $(k_{in})$. The network thus formed has a tree structure, without 
   any loops. Similar to BA model we expect that the main results of our model
   should be robust with respect to the value of $m_o$ used \cite {albert}.

\begin{figure}[t]
\begin{center}
\includegraphics[width=8cm]{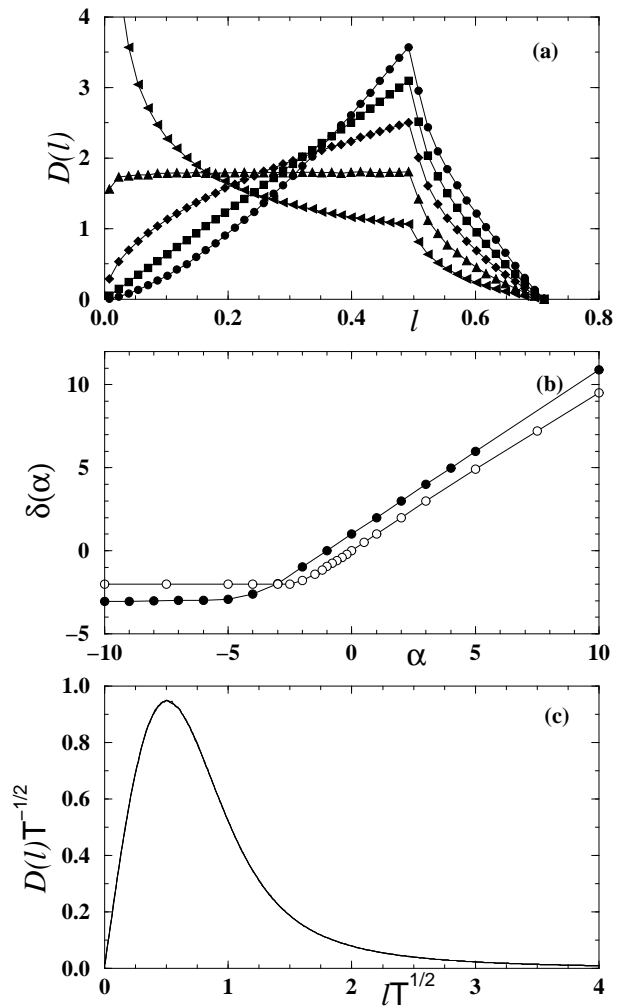}
\end{center}
\caption{
(a) The probability density distribution $D(\ell)$ of the link lengths ${\ell}$ in $d=2$
for five different $\alpha$ values: 1/2 (circle), 0 (square), -1/2 (diamond),
-1 (triangle up), -3/2 (triangle left) for networks of ${\cal T}=2^{12}$
(b) $\delta(\alpha)$ varies linearly with $\alpha$ for $\alpha > -4$ in $d=2$
and for $\alpha > -2$ in $d=1$. The saturation values of $\delta$ are -3 and -2
for $d=2$ (filled circle) and $d=1$ (opaque circle) respectively.
(c) Scaling of the probability density for $\alpha = -\infty$ and for 
${\cal T}=2^{11}, 2^{13}$ and $2^{15}$ in $d=2$.
}
\end{figure}

      The link lengths of this network vary over a wide range. We define 
   $D(\ell)d\ell$ as the
   probability that a randomly selected link has the length between $\ell$ and
   $\ell+d\ell$ and assume a power law distribution $D(\ell) \sim \ell^{\delta}$.
   Since the network with $\alpha=0$ has no length dependence
   and since the nodes are located in random positions in space with uniform probabilities,
   $D(\ell)$ for $\alpha=0$ should depend only on the volume of the spherical
   shell between $\ell$ and $\ell+d\ell$ and $D(\ell) \sim \ell^{d-1}$ in the 
   $d$-dimensional space. For $\alpha \ne 0$, this distribution is 
   modified by the factor $\ell^{\alpha}$ of Eqn. 3 giving 
   \begin {equation}
   D(\ell) \sim \ell^{\alpha+d-1}
   \end{equation}
   which implies $\delta(\alpha)=\alpha+d-1$.
   Therefore at a particular value of $\alpha=\alpha_c=1-d$, $\delta(\alpha)=0$
   and the distribution is uniform in any dimension. $D(\ell)$
   grows with $\ell$ for $\alpha > \alpha_c$ and decays with $\ell$ for
   $\alpha < \alpha_c$. 
   Growing, uniform and decaying distributions are shown in Fig. 2(a)
   for different values of $\alpha$ which confirm $\alpha_c=-1$ in $d=2$.
   Similar calculations in $d=1$ show uniform distribution for $\alpha_c=0$.

   Since we distribute nodes within the unit square and link lengths are measured 
   using the periodic boundary conditions along both the $x$ and $y$ directions, 
   the distance between any two nodes can be at most $\ell_o=2^{-1/2}$. Consequently
   all orientations of links of lengths up to $1/2$ are equally likely. However links 
   of lengths greater than $1/2$ have to be oriented more towards to the diagonal directions
   i.e., $y=\pm x$ lines to properly fit in and therefore their orientations are
   not equally likely. This anisotropic effect is observed in $d$=2 when $D(\ell)$ 
   decays at a faster rate for $1/2 < \ell < \ell_o$ since isotropy of the orientation 
   of these links is lost and therefore they are less probable. In contrast,
   we do not see such a region in one dimension in agreement with the 
   theoretical analysis.

\begin{figure}[t]
\begin{center}
\includegraphics[width=8cm]{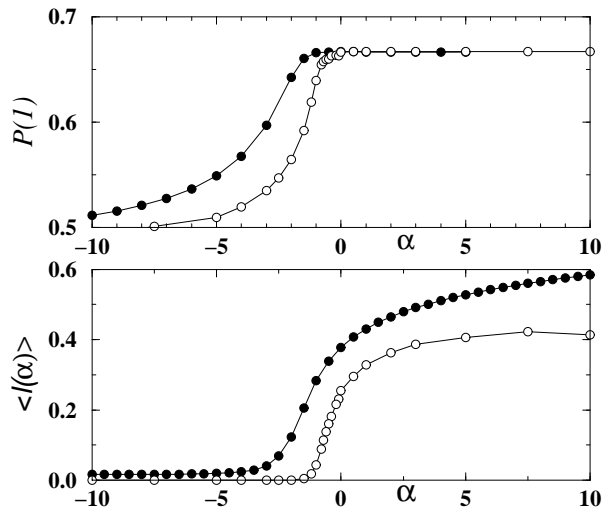}
\end{center}
\caption{
For networks with ${\cal T} = 2^{12}$, variations of 
              (a) the probability $P(1)$ of a node of degree 1
              (b) average length $\ell (\alpha)$ of a link with $\alpha$.
Opaque circles for $d=1$ and the filled circles are for $d=2$.
}
\end{figure}

      However, $\delta(\alpha)$  saturates at a minimum value $\delta_m$ 
   below a certain value of $\alpha$ which is calculated exactly in the limit
   of $\alpha \rightarrow -\infty$. Since, in this limit a new node is always connected
   to its closest neighbour, the probability that the $t+1$-th node has a link
   of length between $\ell$ and $\ell+d\ell$ is 
\begin{equation}
D_t(\ell)d\ell = a\ell^{d-1}t(1-b\ell^d)^{t-1}d\ell
\end{equation}
   where $b\ell^d$ is the volume of a hypersphere in $d$ dimensions and $a=bd$. Therefore
   for a network evolved up to a time ${\cal T}$, the link length distribution of the whole
   network is
\begin{equation}
D(\ell)d\ell = \Sigma_{t=1}^{{\cal T}}D_t(\ell)d\ell 
=a\ell^{d-1}d\ell \Sigma_{t=1}^{{\cal T}} t(1-b\ell^d)^{t-1}.
\end{equation}
   In the limit ${\cal T} \rightarrow \infty$ this series converges for large $\ell$ to 
   $D(\ell) \sim \ell^{-d-1}$ giving $\delta_m=-(d+1)$. However, for small $\ell$, 
   $D(\ell) = c_1 \ell^{d-1} - c_2 \ell^{2d-1}$ ignoring higher order corrections
   where $c_1 \sim O(t^2)$ and $c_2 \sim O(t^3)$ for large $t$. This implies that
   $D(\ell)$ must have a maximum at $\ell \sim {\cal T}^{-1/d}$ for all $d>1$
   as verified below by the scaling analysis. For one dimension however
   no such maximum is expected and the power law decay starts right from the 
   small values of $\ell$.
   Again since $\delta$ varies linearly with $\alpha$ as $\delta(\alpha)=\alpha+d-1$ in general,
   the minimum value $\delta_m$ is attained at $\alpha = -2d$ and remains same for
   smaller values of $\alpha$. 

\begin{figure}[t]
\begin{center}
\includegraphics[width=8cm]{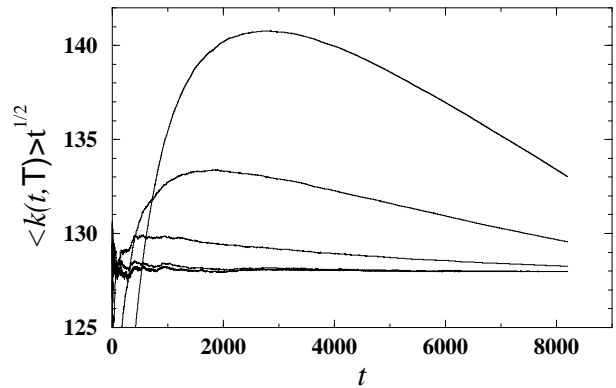}
\end{center}
\caption{
Average degree of the node $\langle k(t,{\cal T}) \rangle$ at time $t$ for a two
dimensional network of duration ${\cal T}=2^{14}$ multiplied by $t^{1/2}$ is plotted
for different $\alpha$.
For $\alpha = 3,2,1,-1$ plots are parallel to the $t$ axis for large $t$.
The curves deviate for $\alpha = -3/2, -2$ and for -3.
}
\end{figure}

      Our numerical findings nicely support these results:
   we check that $\delta$ saturates nearly at $\delta_m = -3$ for 
   $\alpha < -4$ for $d=2$ and $\delta_m = -2$ for $\alpha < -2$ in $d=1$ (Fig. 
   2(b)). We look at the distribution in more detail in $d=2$ for small values of $\ell$.
   On decreasing $\alpha$ from -1 a maximum appears at $\ell \approx 0.01$ (Fig. 2(c)) i.e., as
   $\ell$ increases from zero the distribution grows very rapidly as a power law, reaches 
   a maximum and then decays. For a particular value of $\alpha$, $D(\ell)$ scales
   nicely with the duration ${\cal T}$ of growth as:
   \begin {equation}
   D(\ell) \sim {\cal T}^{1/2} {\cal G}(\ell {\cal T}^{1/2}).
   \end {equation}
   In Fig. 2(c) we plot this collapsed data which fits very well to the following form
   of the scaling function:
   \begin {equation}
   {\cal G}(x)=a'x^{b'}/(x^2+c')^{d'}.
   \end {equation}
   Values of all the constants $a',b',c',d'$ of this expression are dependent on $\alpha$.
   No such maximum in $D(\ell)$ at the small values of $\ell$ is observed in one dimension.

      We studied the cumulative degree distribution $F(k) = \int_k^{\infty} P(k)dk
   \sim k^{1-\gamma}$ and assume the following scaling behaviour:
\begin {equation}
   F(k,{\cal T}) \sim {\cal T}^{\eta}{\cal F}(k/{\cal T}^{\zeta})
\end {equation}
   where the scaling function ${\cal F}(x) \rightarrow 
   x^{1-\gamma}$ when $x << 1$ and ${\cal F}(x) \rightarrow constant$ for $x 
   >> 1$. This implies that $\gamma = 1+\eta/\zeta$. For example for $\alpha=-1$ at
   $d$=2 a good data collapse is obtained for $\eta = 1$ 
   and $\zeta=1/2$ giving $\gamma = 3$ and the same result is obtained for all values of 
   $\alpha > -1$. However for $\alpha < -1$ we see a stretched exponential variation
   $P(k) \sim exp(-k^{\psi(\alpha)})$ and $\psi(\alpha)$ increases
   to 1 i.e., to a pure exponential form at $\alpha = -\infty$.
   It therefore appears that the transition from the stretched exponential 
   to the scale-free behaviour is perhaps taking place at the specific value of $\alpha_c =-1$
   in $d$=2 and in general at $\alpha_c=1-d$ in the $d$ dimension.
   However a similar study in one dimension shows all indications that
   $\alpha_c$ is very likely to be around -0.5, certainly greater than 
   -1 but seems to be smaller than zero. Therefore we conjecture that 
   $\alpha_c=1-d$ though our numerical analysis in one dimention does not fully
   confirm         this prediction.
   At the opposite limit of $\alpha = \infty$ i.e. when each node is connected to its
   farthest neighbour, the degree distribution is found to be exponentially
   decaying and it appears that it happens only at $\alpha = \infty$ since
   even at $\alpha=40$ we found scaling of the distribution.

      We note a few more important properties of this network near $\alpha_c$.
   The first moment of the degree distribution $\langle k \rangle$ is exactly 
   2 since the sum of the degrees of all nodes is 2${\cal T}$ counting each 
   link twice where as the number of nodes is ${\cal T}+1$. As the cut-off of 
   the degree distribution varies as ${\cal T}^{\zeta}$ the $\langle k \rangle$ 
   is defined as $\int_1^{{\cal T}^{\zeta}} kP(k)dk/\int_1^{{\cal T}^{\zeta}} 
   P(k)dk$.  Assuming $\gamma > 2$ the mean is $(\gamma -1)/(\gamma-2)$ in the 
   large ${\cal T}$ limit. This gives $\gamma=3$ correctly. Since each node 
   has exactly one outgoing link, the mean number of incoming links averaged 
   over all nodes must be equal to 1. The fraction of nodes with nonzero incoming 
   links should be 1-$P(1)$. In Fig. 3(a) we plot the variation of $P(1)$ with 
   $\alpha$. This varies from 0.4777 at $\alpha = -\infty$ to very close to 2/3 
   for $\alpha > \alpha_c$. Therefore the mean number of incoming links per 
   node $\langle k_{in}(\alpha) \rangle$ averaged over all nodes with $k_{in} 
   \ne 0$ is ${\cal T}/[(1-P(1)){\cal T}]$=3 for $\alpha > \alpha_c$. This is 
   checked numerically. We also keep track of the variation of average link 
   length $\langle \ell(\alpha) \rangle$ as the characteristic distance of this 
   system which is shown in Fig. 3(b). Since for $\alpha < \alpha_c$ smaller 
   links are more probable, $\langle \ell(\alpha) \rangle$ is very small and 
   approaches zero as ${\cal T}^{-1/2}$. For $\alpha > \alpha_c$, $\langle 
   \ell(\alpha) \rangle$ grows with $\alpha$. Finally the fraction of anisotropic 
   links $f(\ell > 1/2)$ also grows from zero at $\alpha = \alpha_c$.

      The average degree of a node which is introduced at the
   time $t$ to a network which has grown up to a time ${\cal T}$ is 
   denoted by $\langle k(t,{\cal T}) \rangle$ and follows a power law variation with $t$ at large
   $t$ values as: $\langle k(t,{\cal T})\rangle \sim t^{\beta}$.
   For BA network, $\beta=1/2$ is obtained \cite {dorog} and it is connected to the
   degree distribution exponent $\gamma$ by the relation:
   $\beta(\gamma-1)=1$. We have calculated $\langle k(t,{\cal T}) \rangle$ 
   in two dimensions for different values of $\alpha$ and
   in Fig. 4 we plot $\langle k(t,{\cal T}) \rangle t^{1/2}$ vs. $t$ on a linear scale. We observe
   that for all values of $\alpha > -1$ the plot is horizontal implying that 
   $\beta = 1/2$ for this range of $\alpha$ values.

      To summarize, we studied a growing random network where the attachment 
   probability $\pi_i$ to a previous node $i$ depends jointly on the degree of the
   node $k_i$ as well as on the $\alpha$-th power of the link-length $\ell$ as
   $\pi_i \sim k_i \ell^{\alpha}$. By tuning $\alpha$ we find that for
   $\alpha < \alpha_c$, the degree distribution of the
   resulting network is stretched exponential whereas for $\alpha > \alpha_c$ the network
   is scale-free. We also observe that the link length distribution 
   follows a power law: $D(\ell) \sim \ell^{\delta}$ for the whole range of the
   parameter $\alpha$ in contrast to the Waxman's exponential distribution \cite {Waxman}.
   The exponent $\delta$ grows linearly with $\alpha$ for $\alpha \ge -2d$ and
   saturates at $-(d+1)$ for $\alpha < -2d$. Our interesting observation
   is when $\alpha=-d$ the network has the property of real internet network where
   the link length distribution varies inversely with the link length.

      We thank I. Bose, A. Chatterjee, S. Dasgupta and P. A. Sreeram for many 
   useful discussions on network related problems. PS acknowledged financial 
   support from DST grant SP/S2-M11/99.

\leftline {Electronic Address: manna@boson.bose.res.in}
\leftline {parongama@vsnl.net}

\end {document}